\documentclass[prb,showpacs]{revtex4}
\usepackage{graphicx,amssymb,amsmath,color}
\begin{document}
\title{On the effect of weak disorder on the density of states in graphene}
\author{Bal\'azs D\'ora}
\email{dora@pks.mpg.de}
\affiliation{Max-Planck-Institut f\"ur Physik Komplexer Systeme, N\"othnitzer Str. 38, 01187 Dresden, Germany}
\author{Klaus Ziegler}
\affiliation{Institut f\"ur Physik, Universit\"at Augsburg, D-86135 Augsburg, Germany}
\author{Peter Thalmeier}
\affiliation{Max-Planck-Institut f\"ur Chemische Physik fester Stoffe, 01187 Dresden, Germany}

\date{\today}

\begin{abstract}
The effect of weak potential and bond disorder on the density of states of graphene is studied. 
By comparing the self-consistent non-crossing approximation on the honeycomb lattice 
with perturbation theory on the Dirac fermions, we conclude, that the linear density of states of pure graphene 
changes to a non-universal power-law, whose exponent depends on the strength of disorder 
like 1-4$g/\sqrt{3}\pi t^2$, 
with $g$ the variance of the Gaussian disorder, $t$ the hopping integral. This can result in a significant suppression
of the exponent of the density of states in the weak-disorder limit.
We argue, that even a non-linear density of states can result in a conductivity being 
proportional to the number of charge carriers, in accordance with experimental findings.  
\end{abstract}


\maketitle

\section{Introduction}

Graphene is a single sheet of carbon atoms with a honeycomb lattice, exhibiting interesting transport 
properties\cite{novoselov2,geim,sharapov5,mccann,cheianov}. 
These are ultimately connected to the low-energy quasiparticles of graphene, i.e. two-dimensional Dirac fermions. 
Its conductivity depends linearly on the carrier density, and reaches a universal value in the limit of vanishing 
carrier density\cite{novoselov2,geim}. The former has been explained by the presence of charged 
impurities, while the latter does not allow a charged disorder\cite{nomura,castro07}. 
Moreover, in the presence of a magnetic field, the half-integer quantum Hall-effect is explained in 
terms of the unusual Landau quantization and by the existence of zero energy Landau level\cite{geim,sharapov5}.

The density of states in pure graphene is linear around the particle-hole symmetric filling (called the Dirac point), 
and vanishes at the Dirac point. This is a common feature both in the lattice description and in the continuum.
In addition, the lattice model also shows a logarithmic singularity at the hopping energy, which is absent in the 
continuum or Dirac description.

When disorder is present, the emerging picture is blurred. Field-theoretical approaches to related models
(quasiparticles in a d-wave superconductor) 
predict a power-law vanishing with non-universal\cite{ludwig94} or universal\cite{nersesyantsvelik} exponent
or a diverging\cite{ludwig94} density of states, depending on the type and strength of disorder. 
Away from the Dirac point a power-law with positive non-universal exponent is also supported by numerical 
diagonalization of finite size systems\cite{morita,ryu}. At and near the Dirac point the behavior of the
density of states is less clear. Some approaches favour a finite density of states at the Dirac 
point\cite{zieglerhettler}, whereas others predict a vanishing DOS\cite{ludwig94,nersesyantsvelik} or
an infinite DOS\cite{atkinson00}. 
There is some agreement that away from the Dirac point and for weak disorder the DOS behaves like a power
law with positive exponent
\begin{equation}
\rho(E)\sim \rho_0|E|^\gamma \ \ (\gamma>0).
\end{equation}

The purpose of the present paper is to investigate, how a non-universal (therefore disorder dependent) power-law 
exponent (found numerically in Refs. \onlinecite{morita,ryu}) can emerge for weak disorder (compared to the bandwidth), 
and what its physical consequences are.
We determine the exponent 
based on the comparison of the self-consistent non-crossing approximation on the honeycomb lattice and of the 
perturbative treatment of the Dirac Hamiltonian. The exponent decreases linearly with disorder.
Then, using this generally non-linear density of states, we evaluate the conductivity away from the Dirac point by using 
the Einstein relation. We show, that based on the specific form of the diffusion coefficient, this can result in a 
conductivity, 
depending linearly on the carrier concentration\cite{novoselov2}, and in a mobility, decreasing with increasing 
disorder. These are in accord with recent experiment on K adsorbed graphene\cite{chen}.
By varying the K doping time, the impurity strength was controlled. The conductivity away from the Dirac point still 
depends linearly on the charge carrier concentration, but its slope, the mobility decreases steadily with doping time.

Our results apply to other systems with Dirac fermions such as the organic 
conductor\cite{tajima} $\alpha$-(BEDT-TTF)$_2$I$_3$.

\section{Honeycomb dispersion}

We start with the Hamiltonian describing quasiparticles on the honeycomb lattice, given by\cite{semenoff,peresalap}:
\begin{gather}
H_0=h_1\sigma_1+h_2\sigma_2,
\label{hamalap}
\end{gather}
where $\sigma_j$'s are the Pauli matrices, representing the two sublattices. Here, 
\begin{gather}
h_1=-t\sum_{j=1}^3\cos({\bf a}_j{\bf k}),\hspace*{1cm} h_2=-t\sum_{j=1}^3\sin({\bf a}_j{\bf k}),
\end{gather}
with ${\bf a}_1=a(-\sqrt 3/2,1/2)$, ${\bf a}_2=a(0,-1)$ and ${\bf a}_3=a(\sqrt 3/2,1/2)$ pointing towards nearest 
neighbours on the honeycomb lattice, $a$ the lattice constant, $t$ the hopping integral.
The resulting honeycomb dispersion is given by $\pm\sqrt{h_1^2+h_2^2}$, which vanishes at six points in the Brillouin 
zone.
To take scattering into account, we consider the mutual coexistence of both Gaussian potential (on-site) disorder (with 
matrix element $V_{o,r}$, 
satisfying $\langle V_{o,r}\rangle=0$ and variance $\langle V_{o,r}V_{o,r^\prime}\rangle g_0=g_o\delta_{rr^\prime}$) 
and bond disorder in only one direction 
(in addition to the uniform hopping with matrix element $V_{b,r}$, satisfying $\langle V_{b,r}\rangle=0$ and variance 
$\langle V_{b,r}V_{b,r^\prime}\rangle =g_b\delta_{rr^\prime}$), which is 
thought to describe reliably the more complicated case of disorder on all bonds\cite{b1ziegler}.
In graphene, ripples can represent the main source of disorder, and are approximated by random nearest-neighbour hopping 
rates, while potential disorder might only be relevant close to the Dirac point\cite{ostrovsky}. 
The corresponding term in the Hamiltonian is
\begin{equation}
V=V_{o,r}\sigma_0+V_{b,r}\sigma_1,
\end{equation}
which results in $H=H_0+V$.

\begin{figure}[h!]
\centering{\includegraphics[width=4cm,angle=90]{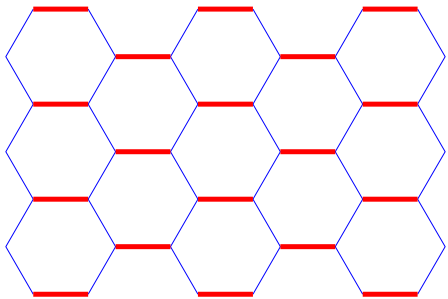}}
\caption{(Color online) A small fragment of the honeycomb lattice is shown. The thick red lines denote the
uni-directional bond disorder, on-site disorder acts on the lattice points.}
\label{honeycomb}
\end{figure}

Without magnetic field, the self-energy for the Green's function, which takes all non-crossing diagrams to every order 
into account (non-crossing approximation, NCA), can be found self-consistently from\cite{vanyolos}
\begin{equation}
\Sigma(i\omega_n)=\dfrac{1}{\dfrac{1-(g_o+g_b)G^2}{[(g_o+g_b)-(g_o-g_b)^2G^2]G}-G},
\label{zerofieldsigma}
\end{equation}
where $i\omega_n$ is the fermionic Matsubara frequency and 
\begin{equation}
G=G_0[i\omega_n-\Sigma(i\omega_n)].
\end{equation}
Here, $G_0$ is the unperturbed local Green's function on the honeycomb lattice given by
\begin{equation}
G_0(z)=\frac{A_c}{(2\pi)^2}\int \frac{zd^2k}{z^2-t^2[4\cos(\sqrt{3}k_x/2)\cos(3k_y/2)+2\cos(\sqrt{3}k_x)+3]},
\end{equation}
where $A_c=3\sqrt{3}a^2/2$ is the area of the unit cell, and the integral runs over the hexagonal Brillouin zone with 
corners given by the 
condition $h_1^2+h_2^2=0$. This can further be brought to a closed form using the results of Ref. 
\onlinecite{horiguchi}.
On the other hand, in the continuum representation, using the Dirac Hamiltonian, the above Green's function simplifies 
to
\begin{equation}
G_0(z)=\frac{2A_c\sigma_0}{(2\pi)^2}\int\frac{d^2k}{z+v(k_x\sigma_1+k_y\sigma_2)}=-\frac{A_c z\sigma_0}{2\pi 
v^2}\ln\left(1-\dfrac{\lambda^2}{z^2}\right),
\label{green0}
\end{equation}
and $v=3ta/2$. The cutoff $\lambda$ can be found by requiring the number of states 
in the Brillouin zone to be preserved in the Dirac case as well. This leads to $\lambda=\sqrt{\pi\sqrt{3}}t$. Another 
possible choice relies on the comparison of the low frequency parts of the Green's function in the lattice and in the 
continuum limit, which reveals the presence of $\ln(\omega/3t)$ terms. This leads to $\lambda=3t$, which coincides with 
the real bandwidth on the lattice. We are going to use 
this form in the following.
The difference of the variances becomes important when calculating the 2nd order correction (in variance) to the 
self-energy.
It is clear from Eq. \eqref{zerofieldsigma}, that the same self-energy is found for pure potential or 
unidirectional bond disorder. 
The effect of their coexistence is the strongest, when they possess the same variance.
From this, the density of states follows as
\begin{equation}
\rho(\omega)=-\frac{1}{\pi}\textmd{Im}G(\omega+i\epsilon)
\end{equation}
with $\epsilon\rightarrow 0^+$. Without disorder, we have the linear density of states 
$\rho(\omega\ll t )=A_c|\omega|/2\pi v^2$.
At zero frequency, in the limit of weak disorder, the self-energy
is obtained as
\begin{equation}
\Sigma(0)=-i\lambda\exp\left(-\dfrac{\pi v^2}{A_c(g_o+g_b)}\right),
\label{sigma0}
\end{equation}
which translates into a residual density of states as
\begin{equation}
\rho(0)=\dfrac{\lambda}{\pi (g_0+g_b)}\exp\left(-\dfrac{\pi v^2}{A_c(g_o+g_b)}\right).
\label{resdosanal}
\end{equation}
From this expression, weak disorder is defined by the condition $g_o+g_b\ll t^2$.
The exponential term indicates the highly non-perturbative nature of density of states
at the Dirac point: all orders of perturbation expansion vanish identically at $\omega=0$.
 
\begin{figure}[h!]
\centering{\includegraphics[width=7cm,height=7cm]{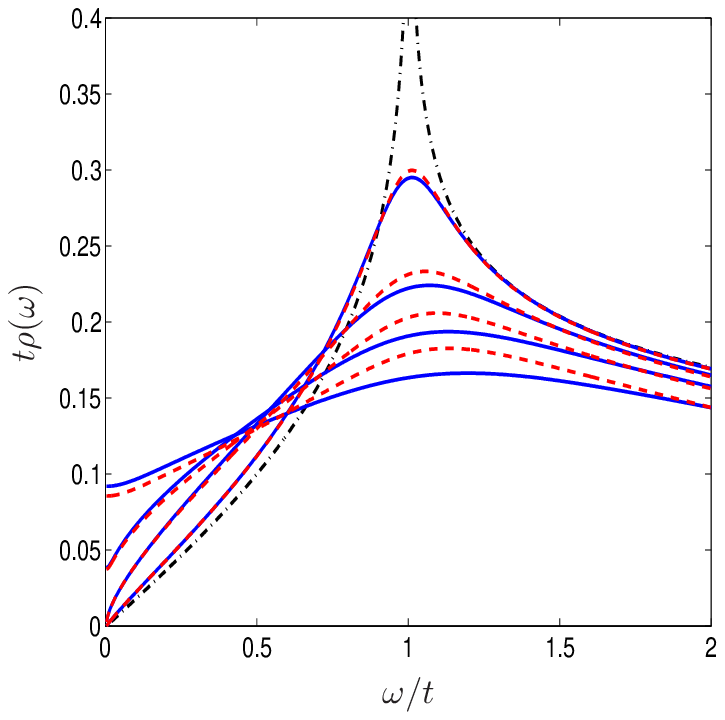}}
\hspace*{1cm}
\centering{\includegraphics[width=7cm,height=7cm]{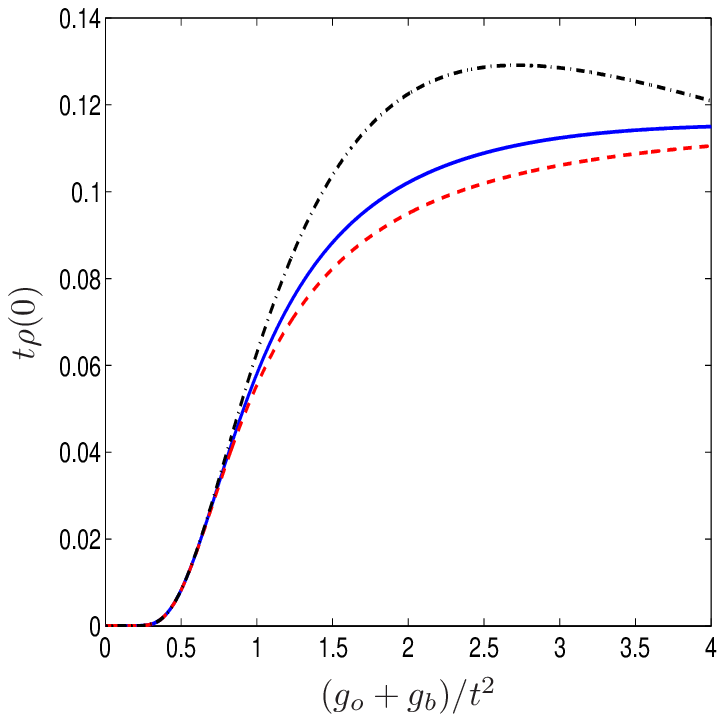}}

\caption{(Color online) The density of states is shown in the left panel for pure on-site ($g_b=0$) or 
unidirectional bond 
($g_o=0$) disorder (solid 
line) for $(g_o+g_b)/t^2=0.1$, 0.4, 0.8 and 1.6 with decreasing DOS at $\omega=t$. The red dashed line represents 
the coexisting bond and unidirectional bond 
disorder with $g_0=g_b$. The black dashed-dotted line denotes the free case with a linear density of states 
at low energies, 
exhibiting a logarithmic divergence at $\omega=t$ in the pure limit. 
The right panel shows the residual density of states for on-site or unidirectional bond disorder 
(blue solid line) and their coexistence with $g_0=g_b$ (red dashed line). 
The black dashed-dotted line denotes the approximate expression, Eq. \eqref{resdosanal}, 
for weak disorder. For $g_o+g_b\leq 0.4t^2$, the residual density of states is negligible.} 
\label{dosmix}
\end{figure}

From Eq. \eqref{zerofieldsigma} it is evident, that the interference of the mutual coexistence of both on-site and bond 
disorder should 
be the most pronounced when $g_o=g_b$. 
The frequency dependence of the density of states on the honeycomb lattice can be obtained by the numerical solution
of the self-consistency equation, Eq. \eqref{zerofieldsigma}, and is shown in Fig. \ref{dosmix}. 
For small frequency and disorder, there is hardly any difference between pure on-site or unidirectional
bond disorder and 
their coexistence. However, at higher energies and disorder strength, they start to deviate from each other. At 
$\omega=t$, the weak logarithmic divergence is washed out with increasing disorder strength. Such features are absent 
from the Dirac  description, which concentrates on the low energy excitations.
Interestingly, for weak disorder, the residual DOS remains suppressed as 
suggested by Eq. \eqref{resdosanal}, but the initial slope in frequency changes.
In order to determine, whether the exponent or its coefficient or both change with disorder, we perform a 
perturbation expansion in disorder strength using the Dirac Hamiltonian to quantify the 
resulting density of states, and compare it to the numerical solution of the self-consistent non-crossing approximation 
using the honeycomb dispersion.

\begin{figure}[h!]
\includegraphics[width=7cm,height=7cm]{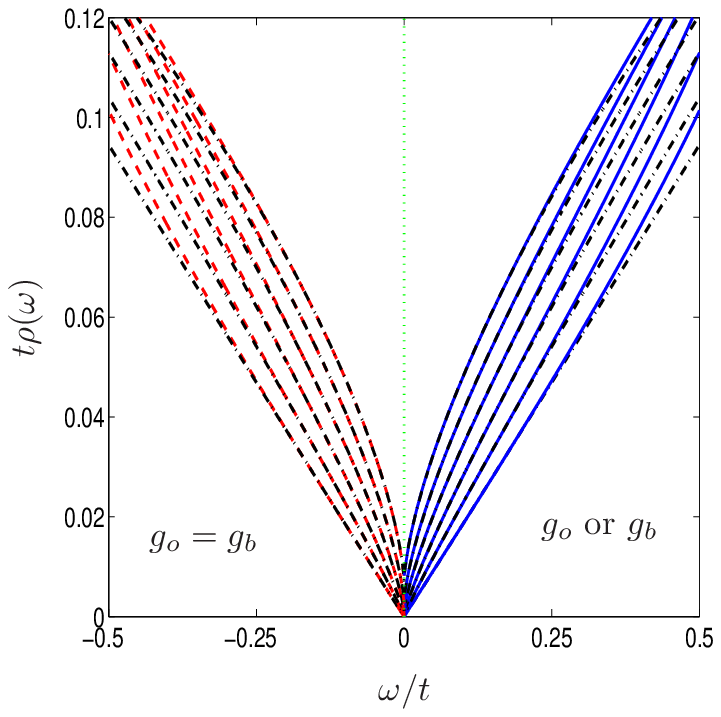}
\hspace*{1cm}
\includegraphics[width=7cm,height=7cm]{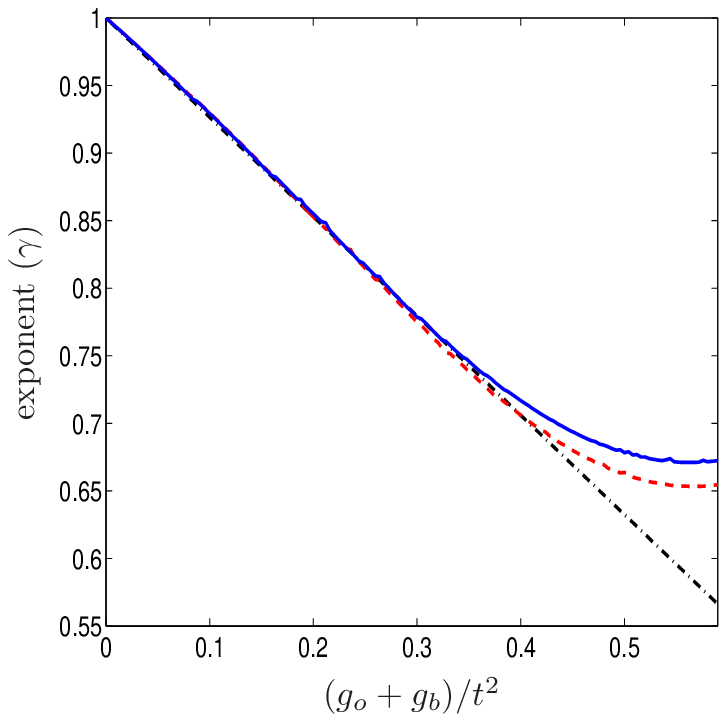}


\caption{(Color online) The low energy density of states is shown in the left panel for $(g_o+g_b)/t^2=0.004$, 
0.1, 0.2, 0.3, 0.4 and 0.5 from bottom to top, for pure on-site ($g_b=0$) or 
unidirectional bond ($g_o=0$)disorder at 
positive energies (blue solid line), and for their coexistence at negative energies at $g_o=g_b$ (red dashed 
line). The black dashed-dotted line denotes the power-law fit as $\rho(\omega)=\rho_0+2\rho_1(\omega/t)^\gamma$.
 The green vertical dotted
line separates the two parts.
The right panel visualizes the exponents as a function of the variance of the disorder, $g_o+g_b$, for on-site or
 bond disorder (blue solid line) and their coexistence ($g_o=g_b$). The black dashed-dotted line denotes the 
result of perturbation theory: $\gamma=1-4(g_o+g_b)/\sqrt 3 \pi t^2$. As is seen, the agreement is excellent in the 
limit 
of weak disorder. Note, that $g$ denotes the variance of the disorder. }
\label{dosmixfit}
\end{figure}

\section{Power-law exponent}

The expansion of the one-particle Green's function in disorder at $z=E+i\epsilon$ leads to
\begin{equation}
G(z)=G_0+G_0VG_0+G_0VG_0VG_0 + ...,
\end{equation}
where $V=V_{o,r}\sigma_0+V_{b,r}\sigma_1$ describes both Gaussian potential and 
unidirectional bond disorder, $G=(z-H_0-V)^{-1}$ and $G_0=(z-H_0)^{-1}$, 
and $H_o=v(k_x\sigma_1+k_y\sigma_2)$ is the Dirac 
Hamiltonian.
After averaging over disorder, we get
\begin{equation}
\langle G_{rr}\rangle=G_{0;rr}+g\sum_{r'}G_{0;rr'}G_{0;r'r'}G_{0;r'r} + ...
\end{equation}
with $g=g_o+g_b$.
The Green's function $G_0$ is translational invariant and reads from Eq. \eqref{green0} at real frequencies as
\begin{equation}
G_{0;rr}=\frac{A_c |E|}{2\pi v^2}\left[\ln\left(\frac{\lambda^2}{E^2}-1\right)-i\pi\right]
\label{green00}
\end{equation}
 This implies
\begin{equation}
\langle G_{rr}\rangle=G_{0;rr}+g(G_0^2)_{rr}G_{0;rr}+ o(g^2)
=G_{0;rr}[1+g(G_0^2)_{rr}]+ o(g^2)
\label{green01}
\end{equation}
Moreover, we have with Eq. \eqref{green00}
\begin{equation}
(G_0^2)_{rr}=\frac{A_c}{(2\pi v^2)^2}\int\frac{d^2k}{[z+v(k_x\sigma_1+k_y\sigma_2)]^2}=
-\frac{\partial G_{0;rr}}{\partial z}
=\frac{A_c}{2\pi v^2}\left[\ln(1-\frac{\lambda}{z^2})-\frac{2\lambda^2}{z^2-\lambda^2}\right]\sigma_0
\approx\frac{A_c}{2\pi v^2}\left[-2\ln\left(\frac{|E|}{\lambda}\right)-i\pi\right]
\end{equation}
for $\lambda\gg |E|\gg\epsilon$. Therefore, we obtain 
\begin{equation}
\langle G_{rr}\rangle= G_{0;rr}
\left[1 - \frac{A_c g}{\pi v^2}\ln\left(\frac{|E|}{\lambda}\right)-\frac{i gA_c}{2 v^2}\right]+ 
o(g^2).
\label{green02}
\end{equation}
From this, the density of states follows as
\begin{equation}
\rho(E)=-\frac{1}{\pi}\textmd{Im}\langle G_{rr}\rangle =\frac{A_c E}{2\pi v^2}
\left[1-\frac{2gA_c}{\pi v^2}\ln\left(\frac{E}{\lambda}\right)\right].
\label{green04}
\end{equation}
If we further assume that (I) the density of states as a function of $E$ 
satisfies a power law (inspired by Refs. \onlinecite{nersesyantsvelik,morita,ryu}) and (II) the disorder strength $g$ is 
small, we can formally consider Eq. \eqref{green04} as the 
lowest order expansion in disorder, and sum it up to a scaling form as
\begin{equation}
 \rho(E)=\frac{A_c\lambda}{2\pi v^2} \left(\frac{E}{\lambda}\right)^{1-(2gA_c/\pi v^2)}.
\label{green03}
\end{equation}
Hence, this suggests that the linear density of states of pure graphene changes into a non-universal 
power-law depending on the strength of the disorder as $\gamma=1-(4g/\pi\sqrt 3 t^2)$. Note, that the exponent does not 
depend on the ambiguous cutoff $\lambda$. 
We mention that Eq. \eqref{green04} might suggest other closed forms than Eq. \eqref{green03}. By using the 
renormalization group procedure to select the most divergent diagrams at a given order $g$ (similarly to parquet 
summation in the Kondo problem), one can sum it up as a geometrical series\cite{efetov,ostrovsky}.
However, the resulting expression contains a singularity around $|\Sigma(0)|$ (Eq. \eqref{sigma0}, playing the role of 
the Kondo temperature here), and is valid 
at high energies compared to $|\Sigma(0)|$ as Eq. \eqref{green04}. To avoid such problems, we use a 
different scaling function, suggested by the results of Refs. \onlinecite{nersesyantsvelik,morita,ryu}.

We compare this expression to the numerical 
solution of the 
self-consistent non-crossing approximation in Fig. 
\ref{dosmixfit}. To extract the exponent, we fit the data with $\rho(E)=\rho_0+2\rho_1(|E|/t)^\gamma$, 
and extract $\rho_{0,1}$ and the exponent $\gamma$.
As can be seen in the left panel, the power-law fits are excellent in an extended frequency window up to 
$t/4$. This suggests that this effect should also be observable experimentally as well. 
The obtained value of $\rho_0$ is negligibly small, as follows from Eq. \eqref{resdosanal}.
From the fits, 
we deduce the exponent and its coefficient, which is shown in the right panel. It agrees well with the 
result of perturbation theory, Eq. \eqref{green03} in the limit of  weak disorder. 
The suppression of the exponent is significant, and can be as big as 30-35\% around $(g_o+g_b)/t^2\sim 0.4$.	
Similar phenomenon has been observed for Dirac fermions on a square lattice in the presence of random 
hopping\cite{morita,ryu}, where disordered systems were studied by exact diagonalization. The decreasing exponent with 
disorder agrees with our results.

\section{Conductivity for non-linear density of states}

Now we turn to the discussion of the conductivity in graphene. A possible starting point is the Einstein
relation\cite{abrikosov}, which states for the conductivity
\begin{equation}
\sigma=e^2\rho D,
\end{equation}
where $\rho$ is the density of states and $D$ is the diffusion coefficient, both at the Fermi energy $E_F$.
Assuming a general power-law density of states, as found above, we have $\rho(E)=\rho_1(E/\lambda)^\gamma$.
In the weak-disorder limit and away from the Dirac point $E=0$, we can safely neglect any tiny residual value.
Moreover, the diffusion coefficient in this case is of the form $D=D_1 E$, which is
validated from the Boltzmann approach in the presence of charged impurities\cite{nomura}.
On the other hand, at the Dirac point there is a exponentially small density of states and
a finite non-zero diffusion coefficient $D\propto g/\rho$ in the presence of uncorrelated bond disorder
\cite{ziegler06}
such that the conductivity is of order 1, in units of $e^2/h$ .
In the following, however, we will concentrate on the regime away from the Dirac point.
Putting these results together, we find
\begin{equation}
\sigma=e^2\rho_1D_1\frac{E_F^{\gamma+1}}{\lambda^\gamma}.
\label{cond1}
\end{equation}
This can be simplified further by noticing that the total number of charge carriers,
participating in electric transport, can be expressed as
\begin{equation}
n=\int_{0}^{E_F}\rho(E)dE=\rho_1\frac{E_F^{\gamma+1}}{(\gamma+1)\lambda^\gamma}.
\end{equation}   
By inserting this back to Eq. \eqref{cond1}, we can read off the conductivity as
\begin{equation}
\sigma=e^2D_1(\gamma+1)n.
\label{cond2}
\end{equation}
From this we can draw several conclusions. First, it predicts that away from the Dirac point, where
our approach predicts a general power-law density of states, the conductivity varies linearly with the
density of charge carriers, in agreement with experiments\cite{novoselov2}.
Second, the mobility of the carriers, which is the coefficient of the $n$ linear term in the conductivity,
behaves as
\begin{equation}
\mu=e\left(2-\frac{4g}{\pi\sqrt 3 t^2}\right)D_1,
\label{mobility}
\end{equation}
where we used our approximate expression for the exponent in the density of states, Eq. \eqref{green03}.
This means, that with increasing disorder ($g$), the mobility decreases steadily, in agreement with recent
experiments on K adsorbed graphene\cite{chen}. There, the graphene sample was doped by K, representing a source 
of charged impurities. Nevertheless, these centers also distort the local electronic environment, and act as bond and 
potential disorder as well. The observed conductivity varied linearly with the carrier 
concentration $n$, similarly to Eq. \eqref{cond2}. Moreover, the mobility (the slope of the $n$ linear term) decreased 
steadily with the doping time (and hence the impurity concentration), which, in our picture, corresponds to a reduction
of the exponent $\gamma$ as well as the mobility, Eq. \eqref{mobility}.

To study the properties close to the Dirac point we have to go beyond the perturbative 
regime.  Then we realize
that the density of states does not vanish at $E=0$. As an approximation we add
a small contribution near the Dirac point
\begin{equation*}
\rho(E)=\rho_0\delta_\eta(E)+\rho_1\left(\frac{E}{\lambda}\right)^\gamma \ ,
\end{equation*}
in form of a soft Dirac Delta function
\begin{equation*}
\delta_\eta(E)=\frac{1}{\pi}\frac{\eta}{E^2+\eta^2} \ \ \ (\eta>0)\ .
\end{equation*}
This implies a particle density $n$ which does not vanish at the 
Dirac point:
\begin{equation}
n(E_F)=\int_0^{E_F}\rho(E)dE\approx \rho_0 + \frac{\rho_1}{(\gamma+1)\lambda^\gamma}
E_F^{\gamma+1} \ .
\label{density}
\end{equation}
Moreover, the diffusion coefficient does neither diverge nor vanish at the Dirac
point\cite{b1ziegler,ziegler06} such that we can assume
\begin{equation*}
D(E)=D_0\delta_\eta(E)+D_1E \ .
\end{equation*}
From the Einstein relation we get the conductivity
which provides an interpolation between a behavior linear in $n$ away
from the Dirac point and a minimal conductivity at the Dirac point:
\begin{equation}
\sigma\sim\frac{e^2}{h} 
\begin{cases}
D_0\rho_0\delta_\eta^2(E_F) & \text{for } E_F\sim0 \\
D_1\rho_1E_F^{\gamma+1}/\lambda^\gamma\sim (1+\gamma)n & \text{for } E_F\gg 0 \\
\end{cases}
 \ .
\label{condfinal}
\end{equation}
This, together with Eq. (\ref{density}), implies for $E_F\gg 0$
the same behavior as in Eq. (\ref{cond2}) with the mobility of Eq. (\ref{mobility}).
The value of the minimal conductivity can be adjusted by choosing the parameter $\eta$ properly.
Therefore Eq. \eqref{condfinal} provides us with a qualitative understanding of the conductivity 
in graphene for 
arbitrary carrier density.

\section{Conclusions}

We have studied the effect of weak on-site and bond disorder on the density of states and conductivity of graphene.
By using the honeycomb dispersion, we determine the self-energy due to disorder in the self-consistent non-crossing 
approximation. The density of states at the Dirac point is filled in for arbitrarily weak disorder. 
We investigate the possibility of observing non-linear density of states away from the Dirac point, motivated by 
numerical studies on disordered Dirac fermionic systems. By comparing the results of non-crossing approximation on the 
honeycomb lattice to perturbation theory in the Dirac case, we 
conclude, that a disorder dependent exponent can account for the evaluated density of states. The exponent decreases 
linearly with the variance for weak impurities. Then, by using the obtained power-law DOS, we evaluate the conductivity 
away from the Dirac point through the Einstein relation. We find, that this causes the conductivity to depend 
linearly on the carries concentration by assuming that the diffusion coefficient is linear in energy\cite{nomura}, 
and the mobility decreases steadily with increasing disorder. These can also be relevant for other systems with Dirac 
fermions\cite{tajima}.

\begin{acknowledgments}
We acknowledge enlighting discussions with A. V\'anyolos.
This work was supported by the Hungarian Scientific Research Fund under grant number OTKA TS049881 and in part by the 
Swedish Research Council.
\end{acknowledgments}

\bibliographystyle{apsrev}
\bibliography{refgraph}

\begin{thebibliography}{24}
\expandafter\ifx\csname natexlab\endcsname\relax\def\natexlab#1{#1}\fi
\expandafter\ifx\csname bibnamefont\endcsname\relax
  \def\bibnamefont#1{#1}\fi
\expandafter\ifx\csname bibfnamefont\endcsname\relax
  \def\bibfnamefont#1{#1}\fi
\expandafter\ifx\csname citenamefont\endcsname\relax
  \def\citenamefont#1{#1}\fi
\expandafter\ifx\csname url\endcsname\relax
  \def\url#1{\texttt{#1}}\fi
\expandafter\ifx\csname urlprefix\endcsname\relax\def\urlprefix{URL }\fi
\providecommand{\bibinfo}[2]{#2}
\providecommand{\eprint}[2][]{\url{#2}}

\bibitem[{\citenamefont{Novoselov et~al.}(2005)\citenamefont{Novoselov, Geim,
  Morozov, Jiang, Katsnelson, Grigorieva, Dubonos, and Firsov}}]{novoselov2}
\bibinfo{author}{\bibfnamefont{K.~S.} \bibnamefont{Novoselov}},
  \bibinfo{author}{\bibfnamefont{A.~K.} \bibnamefont{Geim}},
  \bibinfo{author}{\bibfnamefont{S.~V.} \bibnamefont{Morozov}},
  \bibinfo{author}{\bibfnamefont{D.}~\bibnamefont{Jiang}},
  \bibinfo{author}{\bibfnamefont{M.~I.} \bibnamefont{Katsnelson}},
  \bibinfo{author}{\bibfnamefont{I.~V.} \bibnamefont{Grigorieva}},
  \bibinfo{author}{\bibfnamefont{S.~V.} \bibnamefont{Dubonos}},
  \bibnamefont{and} \bibinfo{author}{\bibfnamefont{A.~A.}
  \bibnamefont{Firsov}}, \bibinfo{journal}{Nature}
  \textbf{\bibinfo{volume}{438}}, \bibinfo{pages}{197} (\bibinfo{year}{2005}).

\bibitem[{\citenamefont{Geim and Novoselov}(2007)}]{geim}
\bibinfo{author}{\bibfnamefont{A.~K.} \bibnamefont{Geim}} \bibnamefont{and}
  \bibinfo{author}{\bibfnamefont{K.~S.} \bibnamefont{Novoselov}},
  \bibinfo{journal}{Nature Materials} \textbf{\bibinfo{volume}{6}},
  \bibinfo{pages}{183} (\bibinfo{year}{2007}).

\bibitem[{\citenamefont{Gusynin and Sharapov}(2006)}]{sharapov5}
\bibinfo{author}{\bibfnamefont{V.~P.} \bibnamefont{Gusynin}} \bibnamefont{and}
  \bibinfo{author}{\bibfnamefont{S.~G.} \bibnamefont{Sharapov}},
  \bibinfo{journal}{Phys. Rev. B} \textbf{\bibinfo{volume}{73}},
  \bibinfo{pages}{245411} (\bibinfo{year}{2006}).

\bibitem[{\citenamefont{McCann et~al.}(2006)\citenamefont{McCann, Kechedzhi,
  Fa{l'}ko, Suzuura, Ando, and Altshuler}}]{mccann}
\bibinfo{author}{\bibfnamefont{E.}~\bibnamefont{McCann}},
  \bibinfo{author}{\bibfnamefont{K.}~\bibnamefont{Kechedzhi}},
  \bibinfo{author}{\bibfnamefont{V.~I.} \bibnamefont{Fa{l'}ko}},
  \bibinfo{author}{\bibfnamefont{H.}~\bibnamefont{Suzuura}},
  \bibinfo{author}{\bibfnamefont{T.}~\bibnamefont{Ando}}, \bibnamefont{and}
  \bibinfo{author}{\bibfnamefont{B.~L.} \bibnamefont{Altshuler}},
  \bibinfo{journal}{Phys. Rev. Lett.} \textbf{\bibinfo{volume}{97}},
  \bibinfo{pages}{146805} (\bibinfo{year}{2006}).

\bibitem[{\citenamefont{Cheianov and Fa{l'}ko}(2006)}]{cheianov}
\bibinfo{author}{\bibfnamefont{V.~V.} \bibnamefont{Cheianov}} \bibnamefont{and}
  \bibinfo{author}{\bibfnamefont{V.~I.} \bibnamefont{Fa{l'}ko}},
  \bibinfo{journal}{Phys. Rev. Lett.} \textbf{\bibinfo{volume}{97}},
  \bibinfo{pages}{226801} (\bibinfo{year}{2006}).

\bibitem[{\citenamefont{Nomura and MacDonald}(2006)}]{nomura}
\bibinfo{author}{\bibfnamefont{K.}~\bibnamefont{Nomura}} \bibnamefont{and}
  \bibinfo{author}{\bibfnamefont{A.~H.} \bibnamefont{MacDonald}},
  \bibinfo{journal}{Phys. Rev. Lett.} \textbf{\bibinfo{volume}{96}},
  \bibinfo{pages}{256602} (\bibinfo{year}{2006}).

\bibitem[{\citenamefont{Neto et~al.}()\citenamefont{Neto, Guinea, Peres,
  Novoselov, and Geim}}]{castro07}
\bibinfo{author}{\bibfnamefont{A.~H.~C.} \bibnamefont{Neto}},
  \bibinfo{author}{\bibfnamefont{F.}~\bibnamefont{Guinea}},
  \bibinfo{author}{\bibfnamefont{N.~M.~R.} \bibnamefont{Peres}},
  \bibinfo{author}{\bibfnamefont{K.~S.} \bibnamefont{Novoselov}},
  \bibnamefont{and} \bibinfo{author}{\bibfnamefont{A.~K.} \bibnamefont{Geim}},
  \bibinfo{note}{arXiv:cond-mat/0709.1163v1}.

\bibitem[{\citenamefont{Ludwig et~al.}(1994)\citenamefont{Ludwig, Fisher,
  Shankar, and Grinstein}}]{ludwig94}
\bibinfo{author}{\bibfnamefont{A.~W.~W.} \bibnamefont{Ludwig}},
  \bibinfo{author}{\bibfnamefont{M.~P.~A.} \bibnamefont{Fisher}},
  \bibinfo{author}{\bibfnamefont{R.}~\bibnamefont{Shankar}}, \bibnamefont{and}
  \bibinfo{author}{\bibfnamefont{G.}~\bibnamefont{Grinstein}},
  \bibinfo{journal}{Phys. Rev. B} \textbf{\bibinfo{volume}{50}},
  \bibinfo{pages}{7526} (\bibinfo{year}{1994}).

\bibitem[{\citenamefont{Nersesyan et~al.}(1994)\citenamefont{Nersesyan,
  Tsvelik, and Wenger}}]{nersesyantsvelik}
\bibinfo{author}{\bibfnamefont{A.~A.} \bibnamefont{Nersesyan}},
  \bibinfo{author}{\bibfnamefont{A.~M.} \bibnamefont{Tsvelik}},
  \bibnamefont{and} \bibinfo{author}{\bibfnamefont{F.}~\bibnamefont{Wenger}},
  \bibinfo{journal}{Phys. Rev. Lett.} \textbf{\bibinfo{volume}{72}},
  \bibinfo{pages}{2628} (\bibinfo{year}{1994}).

\bibitem[{\citenamefont{Morita and Hatsugai}(1997)}]{morita}
\bibinfo{author}{\bibfnamefont{Y.}~\bibnamefont{Morita}} \bibnamefont{and}
  \bibinfo{author}{\bibfnamefont{Y.}~\bibnamefont{Hatsugai}},
  \bibinfo{journal}{Phys. Rev. Lett.} \textbf{\bibinfo{volume}{79}},
  \bibinfo{pages}{3728} (\bibinfo{year}{1997}).

\bibitem[{\citenamefont{Ryu and Hatsugai}(2001)}]{ryu}
\bibinfo{author}{\bibfnamefont{S.}~\bibnamefont{Ryu}} \bibnamefont{and}
  \bibinfo{author}{\bibfnamefont{Y.}~\bibnamefont{Hatsugai}},
  \bibinfo{journal}{Phys. Rev. B} \textbf{\bibinfo{volume}{65}},
  \bibinfo{pages}{033301} (\bibinfo{year}{2001}).

\bibitem[{\citenamefont{Ziegler et~al.}(1996)\citenamefont{Ziegler, Hettler,
  and Hirschfeld}}]{zieglerhettler}
\bibinfo{author}{\bibfnamefont{K.}~\bibnamefont{Ziegler}},
  \bibinfo{author}{\bibfnamefont{M.~H.} \bibnamefont{Hettler}},
  \bibnamefont{and} \bibinfo{author}{\bibfnamefont{P.~J.}
  \bibnamefont{Hirschfeld}}, \bibinfo{journal}{Phys. Rev. Lett.}
  \textbf{\bibinfo{volume}{77}}, \bibinfo{pages}{3013} (\bibinfo{year}{1996}).

\bibitem[{\citenamefont{Atkinson et~al.}(2000)\citenamefont{Atkinson,
  Hirschfeld, MacDonald, and Ziegler}}]{atkinson00}
\bibinfo{author}{\bibfnamefont{W.~A.} \bibnamefont{Atkinson}},
  \bibinfo{author}{\bibfnamefont{P.~J.} \bibnamefont{Hirschfeld}},
  \bibinfo{author}{\bibfnamefont{A.~H.} \bibnamefont{MacDonald}},
  \bibnamefont{and} \bibinfo{author}{\bibfnamefont{K.}~\bibnamefont{Ziegler}},
  \bibinfo{journal}{Phys. Rev. Lett.} \textbf{\bibinfo{volume}{85}},
  \bibinfo{pages}{3926} (\bibinfo{year}{2000}).

\bibitem[{\citenamefont{Chen et~al.}()\citenamefont{Chen, Jang, Fuhrer,
  Williams, and Ishigami}}]{chen}
\bibinfo{author}{\bibfnamefont{J.~H.} \bibnamefont{Chen}},
  \bibinfo{author}{\bibfnamefont{C.}~\bibnamefont{Jang}},
  \bibinfo{author}{\bibfnamefont{M.~S.} \bibnamefont{Fuhrer}},
  \bibinfo{author}{\bibfnamefont{E.~D.} \bibnamefont{Williams}},
  \bibnamefont{and} \bibinfo{author}{\bibfnamefont{M.}~\bibnamefont{Ishigami}},
  \bibinfo{note}{arXiv:0708.2408}.

\bibitem[{\citenamefont{Tajima et~al.}(2007)\citenamefont{Tajima, Sugawara,
  Tamura, Kato, Nishio, and Kajita}}]{tajima}
\bibinfo{author}{\bibfnamefont{N.}~\bibnamefont{Tajima}},
  \bibinfo{author}{\bibfnamefont{S.}~\bibnamefont{Sugawara}},
  \bibinfo{author}{\bibfnamefont{M.}~\bibnamefont{Tamura}},
  \bibinfo{author}{\bibfnamefont{R.}~\bibnamefont{Kato}},
  \bibinfo{author}{\bibfnamefont{Y.}~\bibnamefont{Nishio}}, \bibnamefont{and}
  \bibinfo{author}{\bibfnamefont{K.}~\bibnamefont{Kajita}},
  \bibinfo{journal}{Europhys. Lett.} \textbf{\bibinfo{volume}{80}},
  \bibinfo{pages}{47002} (\bibinfo{year}{2007}).

\bibitem[{\citenamefont{Semenoff}(1984)}]{semenoff}
\bibinfo{author}{\bibfnamefont{G.~W.} \bibnamefont{Semenoff}},
  \bibinfo{journal}{Phys. Rev. Lett.} \textbf{\bibinfo{volume}{53}},
  \bibinfo{pages}{2449} (\bibinfo{year}{1984}).

\bibitem[{\citenamefont{Peres et~al.}(2006)\citenamefont{Peres, Guinea, and
  {Castro Neto}}}]{peresalap}
\bibinfo{author}{\bibfnamefont{N.~M.~R.} \bibnamefont{Peres}},
  \bibinfo{author}{\bibfnamefont{F.}~\bibnamefont{Guinea}}, \bibnamefont{and}
  \bibinfo{author}{\bibfnamefont{A.~H.} \bibnamefont{{Castro Neto}}},
  \bibinfo{journal}{Phys. Rev. B} \textbf{\bibinfo{volume}{73}},
  \bibinfo{pages}{125411} (\bibinfo{year}{2006}).

\bibitem[{\citenamefont{Ziegler}()}]{b1ziegler}
\bibinfo{author}{\bibfnamefont{K.}~\bibnamefont{Ziegler}},
  \bibinfo{note}{arXiv:cond-mat/0703628}.

\bibitem[{\citenamefont{Ostrovsky et~al.}(2006)\citenamefont{Ostrovsky, Gornyi,
  and Mirlin}}]{ostrovsky}
\bibinfo{author}{\bibfnamefont{P.~M.} \bibnamefont{Ostrovsky}},
  \bibinfo{author}{\bibfnamefont{I.~V.} \bibnamefont{Gornyi}},
  \bibnamefont{and} \bibinfo{author}{\bibfnamefont{A.~D.}
  \bibnamefont{Mirlin}}, \bibinfo{journal}{Phys. Rev. B}
  \textbf{\bibinfo{volume}{74}}, \bibinfo{pages}{235443}
  (\bibinfo{year}{2006}).

\bibitem[{\citenamefont{V\'anyolos et~al.}(2007)\citenamefont{V\'anyolos,
  D\'ora, Maki, and Virosztek}}]{vanyolos}
\bibinfo{author}{\bibfnamefont{A.}~\bibnamefont{V\'anyolos}},
  \bibinfo{author}{\bibfnamefont{B.}~\bibnamefont{D\'ora}},
  \bibinfo{author}{\bibfnamefont{K.}~\bibnamefont{Maki}}, \bibnamefont{and}
  \bibinfo{author}{\bibfnamefont{A.}~\bibnamefont{Virosztek}},
  \bibinfo{journal}{New J. Phys.} \textbf{\bibinfo{volume}{9}},
  \bibinfo{pages}{216} (\bibinfo{year}{2007}).

\bibitem[{\citenamefont{Horiguchi}(1972)}]{horiguchi}
\bibinfo{author}{\bibfnamefont{T.}~\bibnamefont{Horiguchi}},
  \bibinfo{journal}{J. Math. Phys.} \textbf{\bibinfo{volume}{13}},
  \bibinfo{pages}{1411} (\bibinfo{year}{1972}).

\bibitem[{\citenamefont{Aleiner and Efetov}(2006)}]{efetov}
\bibinfo{author}{\bibfnamefont{I.~L.} \bibnamefont{Aleiner}} \bibnamefont{and}
  \bibinfo{author}{\bibfnamefont{K.~B.} \bibnamefont{Efetov}},
  \bibinfo{journal}{Phys. Rev. Lett.} p. \bibinfo{pages}{236801}
  (\bibinfo{year}{2006}).

\bibitem[{\citenamefont{{A. A. Abrikosov}}(1998)}]{abrikosov}
\bibinfo{author}{\bibnamefont{{A. A. Abrikosov}}},
  \emph{\bibinfo{title}{Fundamentals of the Theory of Metals}}
  (\bibinfo{publisher}{North-Holland}, \bibinfo{address}{Amsterdam},
  \bibinfo{year}{1998}).

\bibitem[{\citenamefont{Ziegler}(2006)}]{ziegler06}
\bibinfo{author}{\bibfnamefont{K.}~\bibnamefont{Ziegler}},
  \bibinfo{journal}{Phys. Rev. Lett.} \textbf{\bibinfo{volume}{97}},
  \bibinfo{pages}{266802} (\bibinfo{year}{2006}).

\end{thebibliography}
\end{document}